\documentclass{article}
\usepackage{amsfonts}
\usepackage{amsmath}
\usepackage{hyperref}
\usepackage{curves}
\usepackage{enumerate}
\usepackage[utf8]{inputenc}
\usepackage{latexsym}
\usepackage{braket}
\usepackage{mathrsfs}
\usepackage{mathtools}
\usepackage{color}
\usepackage{xcolor}
\usepackage{enumitem}
\usepackage{float}
\restylefloat{table}
\usepackage[a4paper, total={6in, 8in}]{geometry}

\usepackage{cite}
\usepackage[english]{babel}
\usepackage{babelbib}
\usepackage{multirow}
\usepackage{tabularx}
\usepackage{makecell}
\setcellgapes{4pt}
\usepackage[thinlines]{easytable}

\usepackage{colortbl}
\def\bal#1\eal{\begin{align}#1\end{align}}
\usepackage{color,soul}
\usepackage{authblk}
\newcommand{\bsub}{\begin{subequations}}
\newcommand{\esub}{\end{subequations}}
\def\bal#1\eal{\begin{align}#1\end{align}}

\newcommand{\gra}{{\alpha}} \newcommand{\grb}{{\beta}}  
  \newcommand{\grh}{{\eta}} 
  \newcommand{\grl}{{\lambda}} 
\newcommand{\grn}{{\nu}} \newcommand{\grj}{{\xi}}  \newcommand{\grp}{{\pi}}
\newcommand{\grr}{{\rho}} \newcommand{\grs}{{\sigma}}

  \newcommand{\grL}{{\Lambda}}

\date{}
\begin{document}
\author[1]{T. Pailas\thanks{teopailas879@hotmail.com}}
\author[2]{N. Dimakis\thanks{nsdimakis@scu.edu.cn; nsdimakis@gmail.com}}
\author[3,4]{Andronikos Paliathanasis\thanks{anpaliat@phys.uoa.gr}}
\author[1]{Petros A. Terzis\thanks{pterzis@phys.uoa.gr}}
\author[1]{T. Christodoulakis\thanks{tchris@phys.uoa.gr}}
\affil[1]{Department of Nuclear and Particle Physics, Faculty of Physics, National and Kapodistrian University of Athens, Athens 15784, Greece}
\affil[2]{Center for Theoretical Physics, College of Physics, Sichuan University, Chengdu 610064, China}
\affil[3]{Institute of Systems Science, Durban University of Technology, PO Box 1334, Durban 4000, South Africa}
\affil[4]{Instituto de Ciencias F\'{\i}sicas y Matem\'{a}ticas, Universidad Austral de Chile, Valdivia, 5090000, Chile}

\title{\textbf{Infinite dimensional symmetry groups of the Friedmann equations}}
\maketitle

\begin{abstract}
We find the symmetry generators for the Friedman equations emanating from a perfect fluid source, in the presence of a cosmological constant term. The
relevant dynamics is seen to be  governed by two coupled, first order ordinary differential equations, the  continuity and the quadratic constraint equation. Arbitrary functions
appear in the components of the symmetry vector, indicating the infinity of the group. When the equation of state is considered as arbitrary but ab initio given,  previously known results are recovered and/or generalized. When the pressure is considered among the  dynamical variables, solutions for models with different equations of state are mapped to each other; thus enabling the presentation of solutions to models with complicated equations of state starting from simple known cases.
\end{abstract}

\numberwithin{equation}{section}

\section{Introduction}

An interesting discussion on the appearance of symmetry in nature and in
various motifs of art is presented by H. Weyl in \cite{Weyl1}. A symmetric
structure usually represents an economical process in Nature and an
aesthetically pleasing object to human mind. In Art, usually the symmetry is
an expression of invariance under some form of translation, rotation or
reflection. The (Ancient) Greek word for Art, \textquotedblleft $\tau \acute{%
\varepsilon}\chi \nu \eta $\textquotedblright $,$ can be translated today as
Science. In the latter, the concept of symmetry is appealing for it leads to a
characterization of the system under consideration. The invariance
associated with the symmetry produces a simplification which makes
understanding easier.

The concept of symmetry played an important role in the development of
physical theory. Kepler attempted to impose the idea of symmetry for the
description of the motion of the planets. Furthermore, Newton's laws of
mechanics embodied symmetry principles, such are the Galilean invariance or
the principle of equivalence of inertial frames \cite{Djcross}. Another example is
Maxwell's equations which describe the electromagnetic theory: while they possess the
Lorentz invariance and the gauge invariance, these symmetries were
explicitly developed a few decades after Maxwell's work \cite{max1}.

The first systematic approach for the description of symmetry in modern
science was constructed by Sophus Lie towards the end of 19th century \cite%
{lie1,lie2,lie3}. The novelty of Lie's work was to consider the infinitesimal
representations of the finite transformations induced on a manifold by continuous groups; thereby
moving from the group to a local algebraic representation in terms of vector fields on the tangent
space of the manifold. The study of the invariance properties of various geometrical objects is
 then performed using the Lie Derivative with respect to the afore mentioned fields (generators).
This resulted in a linearization of all equations describing the symmetry conditions. The original work of Lie was
motivated by geometric considerations; he commenced with point
transformations, and so point symmetries, and then extended his work by use of
contact transformations \cite{lie4}.

Emmy Noether in her revolutionary work on the invariance of the Action Integral
of the Calculus of Variations under infinitesimal transformations \cite%
{Noether18} introduced routine dependence of the transformation on the
derivatives, without the requirement that the transformation be contact, and thus
the use of generalised transformations has become well-established as a tool
in the study of differential equations, particularly partial differential
equations. In addition, Noether's work provides a systematic method for the
construction of conservation laws for dynamical systems which follow from a
variation principle. In particular there is a direct relation of the
infinitesimal transformations which leave form invariant the Action Integral with
conservation laws for the equations of motion; for a recent review and
discussion on Noether's work we refer the reader to \cite{non11}.

Since Lie's theory provides a systematic way for the treatment of
nonlinear differential equations, by determining exact and analytic solutions
or conservation laws, it plays a significant role in the development of the
fundamental modern physics, such as Analytic Mechanics, Quantum Mechanics,
General Relativity and Cosmology \cite{sy1,sy2,sy3,sy4,sy5}.

In gravitational physics, specifically in General Relativity, the context
of symmetry is essential for the determination of exact solutions to the
Einstein field equations \cite{sy6}. The assumption of existence of a
collineation, such as an isometry, affects the Einstein tensor so that to simplify
the corresponding nonlinear Einstein field equations. For instance, in Bianchi models the gravitational
field equations reduce from partial differential equations to ordinary
differential equations because of the existence of the three-dimensional
isometry group for the metric tensor. Indeed, there are exact solutions for
the Einstein field equations without any isometry \cite{szbook}; however
other kind of symmetries can exist for these spacetimes \cite{sz11}.

Other types of symmetries have also been used before  in order to
study and solve the Einstein field equations: The infinitesimal
transformations of the phase space of the dynamical variables for the field
equations appear in the literature \cite%
{ns03,ns04,ns05,ns13,ns14,ns15,tch10,tch20,bas1,bas2,ns16,bas5}, while in
the recent review \cite{ss01} the various approaches have been categorized and a
discussion is given for the geometric character of the symmetries of the
field equations. The invariants which follow from the latter analysis have
been used for the quantization process in quantum cosmology and the solution of the
corresponding Wheeler-DeWitt equation \cite{ch01}.

A group of infinitesimal transformations which leave invariant a given
dynamical system can be used to determine invariants and conservation laws. The later are applied to simplify the dynamical system and reduce it
to a known integrable class or to write the differential equations
in an algebraic form. Moreover, if a solution for a given dynamical system is
known, then symmetries can be applied to construct new solutions. For
instance, the Gasperini-Veneziano duality property of the dilaton field \cite%
{dd00}, in a spatially flat Friedmann-Lema\^{\i}tre-Robertson-Walker \
(FLRW) background space, is a discrete symmetry which is directly related to
a continuous symmetry \cite{dd11}. Specifically the $O\left( d,d\right) $
invariance of the two-dimensional linear system of Lorentzian signature can
be transformed to the Gasperini-Veneziano duality invariance \cite{dd22}.

Maybe the first formal study on Lie-point symmetries of cosmological equations can be found in \cite{Szy}. In \cite{Chimento} a group of point transformations under which the Einstein field
equations are invariant have been applied for the construction of new
solutions from known ones in the case of spatially flat FLRW spacetime.
While a family of transformations which relate various inflationary
behaviours in the presence of a scalar field were discussed in \cite{newold}.
An alternative reconstruction method of inflationary solutions from other
known solutions was studied in \cite{jdband}. In particular, in the case of
spatially flat FLRW a map was found  which transforms solutions into
solutions, while the different physical models result into equivalent
systems which are characterized by the admitted $SL\left( 3,R\right) $ Lie
algebra as invariant infinitesimal transformations. Recently, in \cite%
{Faraoni}, there appeared a new point transformation which keeps invariant the cosmological
field equations of a spatially flat FLRW background space where the matter
source is that of an ideal gas.

In this work we consider the gravitational field equations of FLRW
spacetime, without  imposing the vanishing of spatial curvature,
while for the matter source we consider the existence of a cosmological
constant and of a perfect fluid with an arbitrary equation of state. We find three different infinite families of
symmetries which leave invariant the field equations, when a corresponding
constraint equation is satisfied by the unknown functions appearing in the generators .
We recover and generalize previous results from the literature while we
demonstrate how the new results can be used to determine new solutions from old.

In particular we show how the symmetry generators found can be used to connect models with different
equations of state and thus to collectively describe the main epochs of the
cosmological evolution, starting from the early inflation era, passing to the
radiation and then to matter eras; these solutions can also be used to
describe the late-time acceleration phase of the universe. Moreover,
starting from the exact solution of a massless scalar field - which is
described by a stiff fluid- we show that under a specific transformation,
the exact solution which correspond to a modified Chaplygin gas model is
recovered. This later model is known to be  describable by a
quintessence scalar field with non-zero potential. Last but not least we
show how an exact solution of a parametric dark energy model can be
constructed by starting from the matter dominated era. \ The plan of the paper
is as follows.

In Section 2, we define the gravitational model of our consideration,
we discuss the concept of symmetry in the resulting ordinary differential
equations, and we perform a detailed derivation of the symmetry vectors in
the most generic scenario. The generators of the transformations which leave
invariant the field equations are determined. In Section 3, we recover
previous results of the literature, which can be seen as special cases of
our analysis. Our general results are applied in Section 4. Finally in
Section 5, we discuss our results and draw our conclusions.

\section{Cosmological field equations and their symmetries}

We adopt the Friedmann–Lema\^{i}tre–Robertson–Walker line element
\begin{equation}\label{lineel}
  ds^2 = - N(t)^2 dt^2 + a(t)^2 \left[\frac{dr^2}{1-k r^2} + r^2 \left(d\theta^2 + \sin^2\theta d\varphi^2 \right) \right]
\end{equation}
and consider a perfect fluid with energy density $\rho$ and pressure $P$. The energy momentum tensor is
\begin{equation}\label{emom}
  T_{\mu\nu} = (\rho+P) u_\mu u_\nu + P g_{\mu\nu},
\end{equation}
where $u_\mu = (N,0,0,0)$ is the comoving velocity, $u^\mu u_\mu=-1$.

The system to be solved is composed by Einstein's field equations
\begin{equation}\label{ein}
  R_{\mu\nu} - \frac{1}{2} g_{\mu\nu} R + \Lambda g_{\mu\nu} = 8\pi T_{\mu\nu} ,
\end{equation}
together with the continuity equation $\nabla_\mu T^{\mu\nu} =0$ (we adopt the units $G=c=1$). The latter, together with the temporal component of \eqref{ein} are first order ordinary differential equations which can be written as
\begin{align} \label{eq1}
  E_1 & := \Lambda -\frac{3 k}{a^2} + 8\pi \rho - 3\left(\frac{\dot{a}}{N a} \right)^2 =0 \\ \label{eq2}
  E_2 & :=  3\frac{\dot{a}}{a} \left(P + \rho \right) + \dot{\rho}=0
\end{align}
The spatial components of \eqref{ein} comprise the unique (due to the high symmetry) equation
\begin{equation}\label{eq3}
  E_3 := \Lambda -\frac{k}{a^2} -8 \pi  P - 3 \left(\frac{\dot{a}}{N a} \right)^2 - \frac{2}{N} \frac{d}{dt} \left( \frac{\dot{a}}{N a}\right) =0
\end{equation}
which, however, does not constitute an independent relation, i.e. it can be obtained by use of \eqref{eq1}, \eqref{eq2} and the time derivative of the first.

Since the system is described by purely first order equations, namely \eqref{eq1} and \eqref{eq2}, we expect it to admit an infinite number of Lie-point symmetries \cite{Stephani,Olver}. We briefly review the basic aspects of obtaining the Lie-point symmetries of a system of  differential equations; for the interested reader we refer to the textbooks \cite{Stephani,Olver}. Assume a set of ordinary differential equations of the form $E_I(t,x,\dot{x},\ddot{x})=0$ like the ones we have here\footnote{For the needs of the general context we assume second order of ordinary differential equations, even though for our main result we need only consider first order ones: $E_1=0$ and $E_2=0$.}.

A vector
\begin{equation}\label{Xvec}
  X= \chi(t,x) \frac{\partial}{\partial t} + \xi^i (t,x) \frac{\partial}{\partial x^i}
\end{equation}
in the space of the dependent and independent variables, $x$ and $t$ respectively, is called a Lie-point symmetry of the system  $E_I=0$ if it satisfies the infinitesimal criterion of invariance
\begin{equation}\label{critinv}
  \mathrm{pr}^2 X (E_I) =0 ,\quad \mathrm{mod} \quad E_J=0, \quad \text{for all } I,J .
\end{equation}
The $\mathrm{pr}^2 X$ is the second (due to considering up to second order equations) prolongation of the vector $X$, i.e. an extension of the original $X$ vector to the space of the derivatives $\dot{x}$ and $\ddot{x}$ and it is given by the formula:
\begin{equation}\label{pr2}
  \mathrm{pr}^2 X = X + \left( \frac{d \xi^i}{dt} - \dot{x}^i \frac{d\chi}{dt} \right) \frac{\partial}{\partial \dot{x}^i} + \left[\frac{d}{dt}\left( \frac{d \xi^i}{dt} - \dot{x}^i \frac{d\chi}{dt} \right)- \ddot{x} \frac{d\chi}{dt}\right] \frac{\partial}{\partial \ddot{x}^i}.
\end{equation}
The extended coefficients of $\mathrm{pr}^2 X$ in the space of $(\dot{x},\ddot{x})$ are such so that it is guaranteed that the transformation law maps functions to functions \cite{Ibra}: Given a generator $X$ that sets a transformation law in the space of $(t,x)$ and the fact that the $x$ are functions of $t$ at the level of the equations, the $\dot{x}=\frac{d x}{d t}$ and $\ddot{x}= \frac{d^2 x}{dt^2}$ obviously cannot transform in a random manner. The correct  transformation law for the derivatives is assured by the formula \eqref{pr2}.

Let us now proceed into applying this general theory onto the Friedmann equations. We are going to deal with two distinct situations: In the first we assume some generic, but supposedly specific equation of state during the transformation. In the second case we treat the pressure as a dynamical variable and we allow it to transform together with the rest of the variables, thus enabling connections between models described by different equations of state.

\subsection{Invariance that preserves the equation of state}

We begin our study by considering a fixed equation of state for $P$ in \eqref{eq1}-\eqref{eq3}. This means that we do not allow for $P$ to variate under the transformation. Thus, our dependent variables are $x=(N,a,\rho)$ and the generator has the general form
\begin{equation}\label{Xvec2}
  X= \chi(t,N,a,\rho) \frac{\partial}{\partial t} + \xi^1(t,N,a,\rho) \frac{\partial}{\partial N} + \xi^2 (t,N,a,\rho)\frac{\partial}{\partial a} + \xi^3 (t,N,a,\rho) \frac{\partial}{\partial \rho} .
\end{equation}

As we mentioned our system is completely characterized by the first order equations $E_1=0$ and $E_2=0$. The symmetry condition \eqref{critinv} states that the prolongation of the vector has to annihilate the equations modulo the equations themselves. We realize it in the following manner: we demand that the action of the prolongation vector returns multiples of the equations, which of course are bound to be zero and thus satisfy \eqref{critinv}. In other words we require
\begin{subequations} \label{symeq}
\begin{align}
  & \mathrm{pr}^2 X (E_1) = \left( \sigma_{1} \dot{a} + \sigma_{2} \dot{\rho} + \sigma_{3} \right) E_1 + \left(\sigma_4 \dot{a}^2 + \sigma_5 \dot{a} + \sigma_6 \right) E_2 \\
  &\mathrm{pr}^2 X (E_2) = \left(\sigma_7 \dot{a} + \sigma_8 \dot{\rho}+ \sigma_9 \right)E_1 + \left(\sigma_{10} \dot{a}^2 + \sigma_{11} \dot{a} +\sigma_{12} \dot{\rho} +\sigma_{13} \dot{N} + \sigma_{14}\right) E_2,
\end{align}
\end{subequations}
where all the $\sigma_i$, $i=1,...,14$ are functions of $t$, $a$, $\rho$ and $N$. The specific dependence on the derivatives $\dot{a}$, $\dot{\rho}$ and $\dot{N}$ in the right hand side of \eqref{symeq} has been chosen so that it is in accordance with the terms produced by the left hand side, hence avoiding any trivial multipliers $\sigma_i$. Since the $\chi$ and $\xi$ functions of the vector \eqref{Xvec2} have no dependence on derivatives (we search for point symmetries), the coefficients of the terms involving derivatives in \eqref{symeq} must be separately set equal to zero. This forms a set of partial differential equations for $\chi$, $\xi$'s and algebraic in the $\sigma_i$ to be solved.

Refraining from presenting the specific solutions for the $\sigma_i$ - since the latter are multipliers of zeros and are not important in our considerations - the general symmetry vector satisfying \eqref{symeq} can be of the following three general categories:
\begin{enumerate}
  \item The vector
    \begin{equation}\label{X1}
      X_1 = \chi_1(t) \chi_2(a,\rho) \frac{\partial}{\partial t} - N \dot{\chi}_1(t) \chi_2(a,\rho) \frac{\partial}{\partial N},
    \end{equation}
    with $\chi_1(t)$ an arbitrary function and $\chi_2(a,\rho)$ satisfying the condition
    \begin{equation}\label{cond1}
      a \frac{\partial \chi_{2}}{\partial a} -3 (P(a,\rho )+\rho ) \frac{\partial \chi_{2}}{\partial \rho} =0 .
    \end{equation}
    The special solution $\chi_2=$constant results in the parametrization invariance generator
    \begin{equation}\label{Xpar}
    X_{par} = \chi_1(t) \frac{\partial}{\partial t} - N \dot{\chi}_1(t) \frac{\partial}{\partial N} .
  \end{equation}
  It describes the well known property of \eqref{eq1},\eqref{eq2} and \eqref{eq3} being invariant under arbitrary changes in the time variable; a symmetry that is present in all cosmological systems \cite{tchris}.
  \item The second case we distinguish is characterized by
  \begin{equation}\label{X2}
      X_2 = \frac{4 \pi  a^2 N f(a,\rho )}{3k -a^2 (\Lambda +8 \pi  \rho )}\frac{\partial}{\partial N} + f(a,\rho)\frac{\partial}{\partial \rho},
    \end{equation}
    under the condition that $f(a,\rho )$ satisfies
    \begin{equation}\label{cond2}
      3 f(a,\rho ) \left(\frac{\partial P}{\partial \rho}+1\right)-3 \frac{\partial f}{\partial \rho} (P(a,\rho )+\rho )+a \frac{\partial f}{\partial a} =0 .
    \end{equation}
  \item Finally we have the symmetry generator
    \begin{equation}\label{X3}
      \begin{split}
        X_3 & =  N \left[ \frac{\partial \xi}{\partial a}  - \frac{3}{a}(P(a,\rho )+\rho ) \frac{\partial \xi}{\partial \rho}
       + \frac{a  \left(12 \pi  P(a,\rho )+4 \pi  \rho -\Lambda \right)}{a^2 (\Lambda +8 \pi  \rho )-3 k} \xi(a,\rho ) \right]\frac{\partial}{\partial N} \\
       & + \xi(a,\rho) \frac{\partial}{\partial a} -\frac{3  (P(a,\rho )+\rho )}{a}\xi(a,\rho ) \frac{\partial}{\partial \rho},
       \end{split}
    \end{equation}
    with $\xi(a,\rho)$ an arbitrary function. Unlike the $\chi_2(a,\rho)$ and $f(a,\rho)$ we introduced previously, the $\xi(a,\rho)$ here does not depend on the equation of state $P=P(a,\rho)$.
\end{enumerate}
By enforcing the resulting generators on the third equation we derive
\begin{equation}\label{symeqb}
  \mathrm{pr}^2 X_i (E_3) = \Sigma^j E_j , \quad i,j=1,2,3,
\end{equation}
where the $\Sigma^i$ are multiplying functions. Thus, the resulting vector fields are automatically symmetries of the third equation since they satisfy condition \eqref{critinv}.

We thus have three distinct infinite dimensional groups. It can be easily checked, with the help of conditions \eqref{cond1} and \eqref{cond2}, that the subalgebras spanned by each of the $X_1$, $X_2$ and $X_3$ vectors separately are closed. For example, the commutator of two type $X_1$ vectors produces again a type $X_1$ vector, etc. In what regards cross commutation relations, it can be shown that $[X_1,X_2]$ is an $X_1$ vector, $[X_2,X_3]$ is a type $X_3$ vector, while $[X_1,X_3]=0$ by virtue of \eqref{cond2}.

Especially in what regards the vector $X_3$, involving the free function $\xi(a,\rho)$, we can directly see why it leaves invariant the continuity equation $E_2=0$ of \eqref{eq2}. The  latter is sensitive only in changes in $a$ and $\rho$.\footnote{Since a change in $t$ for example transforms $\dot{a}$ and $\dot{\rho}$ in the same manner.} By calculating the integral curve of $X_3$ in the $(a,\rho)$ surface we obtain
\begin{equation}
  \frac{da}{\xi} = \frac{d\rho}{-\frac{3  (P(a,\rho )+\rho )}{a}\xi} \Rightarrow \frac{d\rho}{d a} = -\frac{3  (P(a,\rho )+\rho )}{a},
\end{equation}
but this is none other than equation \eqref{eq2} itself.

In the above setting we considered $P=P(a,\rho)$. However, the process can be generalized to include an equation of state that also incorporates the Hubble function. To this end we may assume that we have a pressure of the form $P=P(a,\rho,\frac{\dot{a}}{N})$. The combination $\frac{\dot{a}}{N}$ is chosen  due to being a scalar under time transformations and since the Hubble function in a generic time gauge reads: $H=\frac{1}{N}\frac{\dot{a}}{a}$. Let us assume for simplicity that the equation involves just some power of this term, i.e.
\begin{equation}\label{presH}
  P = P_1(a,\rho) + P_2(a,\rho) \left( \frac{\dot{a}}{N} \right)^\mu .
\end{equation}

With such an expression, by considering $P_2 \propto a^{-\mu}$, we can create a dependence on the $\mu$-th power of the Hubble function. But, irrespectively of the latter, and avoiding the meticulous details, it can be verified that the following generators are symmetries of the Friedmann equations when expression \eqref{presH} is considered:
\begin{enumerate}
  \item The generator $X_1$ of \eqref{X1} under the condition
    \begin{equation} \label{cond1b}
      a \frac{\partial \chi_2}{\partial a} - 3 \left[\rho+ P_1(a,\rho )+ 3 P_2 (a,\rho ) \Sigma(a,\rho)^{\mu /2} \right]  \frac{\partial \chi_2}{\partial \rho}=0 .
    \end{equation}
  \item The $X_2$ of \eqref{X2} whenever
    \begin{equation} \label{cond2b}
      \begin{split}
        & 3 f(a,\rho ) \left[\frac{\partial P_1}{\partial \rho}+1 + \Sigma(a,\rho )^{\mu /2}  \frac{\partial P_2}{\partial \rho} + \frac{4\mu \pi}{3}   a^2  \Sigma(a,\rho )^{\frac{\mu -2}{2}} P_2(a,\rho ) \right] \\
        & -3 \left[P_1(a,\rho )+ \rho + P_2(a,\rho ) \Sigma(a,\rho )^{\mu /2} \right] \frac{\partial f}{\partial \rho} +\frac{\partial f}{\partial a} =0 .
      \end{split}
  \end{equation}
   The comparison to the previous cases is direct, since setting $P_2=0$ in \eqref{cond1b} and \eqref{cond2b} leads to \eqref{cond1} and \eqref{cond2} respectively.
   \item Finally we have the symmetry generator
   \begin{equation}\label{X3b}
     \begin{split}
       X_3 = & N \left[ \frac{\partial \xi}{\partial a} -\frac{3 \left(P_1 + P_2 \Sigma(a,\rho )^{\mu /2} + \rho \right)}{a} \frac{\partial \xi}{\partial \rho} +  \frac{a \left(12 \pi  \left(P_1+ P_2 \Sigma(a,\rho )^{\mu /2} \right) + 4 \pi  \rho - \Lambda  \right)}{3 \Sigma(a,\rho )} \xi(a,\rho)  \right] \frac{\partial}{\partial N} \\
        & + \xi(a,\rho) \frac{\partial}{\partial a} -\frac{3 \left(P_1+P_2 \Sigma(a,\rho )^{\mu /2}+ \rho \right)}{a} \xi(a,\rho) \frac{\partial}{\partial \rho}
     \end{split}
   \end{equation}
   with $\xi(a,\rho)$ an arbitrary function and where  for simplification we have set $\Sigma(a,\rho ) =  \frac{1}{3} a^2 (\Lambda +8 \pi  \rho ) - k$ in all the previous relations. Obviously setting $P_2=0$ again we return to the previous case and specifically to generator \eqref{X3}.
\end{enumerate}
It is easy to verify that all of the above generators satisfy $\mathrm{pr}^2X_i(E_j)=0$, $i,j=1,2,3$, whenever $E_1=0$, $E_2=0$ and $E_3=0$. Once more, the invariance of the two first leads directly to the invariance of the third.

\subsection{Invariance connecting different models}

Up to now we discussed invariance of our equations for a given (even though not explicitly specified) equation of state. It is more interesting however to allow for $P$ to be dynamical and thus change under a symmetry transformation. In this manner we can use a symmetry vector in order to map known solutions corresponding to simple equations of state, to other theories involving more complicated expressions for the latter.

To this end let us consider the following possible generator
\begin{equation}\label{Xvec3}
  \begin{split}
    X=  \chi(t,N,a,\rho,P) \frac{\partial}{\partial t} + \xi^1(t,N,a,\rho,P) \frac{\partial}{\partial N} + \xi^2(t,N,a,\rho,P) \frac{\partial}{\partial a}  & + \xi^3 (t,N,a,\rho,P) \frac{\partial}{\partial \rho}  \\
    & + \xi^4 (t,N,a,\rho,P) \frac{\partial}{\partial P} .
  \end{split}
\end{equation}
So we extended the vector by assuming a $\partial_P$ component, while all the coefficients may additionally depend on $P$. The symmetry condition \eqref{critinv} is now expressed as
\begin{subequations} \label{symeq2}
\begin{align}
  & \mathrm{pr}^2 X (E_1) = \left( \sigma_{1} \dot{a} + \sigma_{2} \dot{\rho} + \sigma_{3} \right) E_1 + \left(\sigma_4 \dot{a}^2 + \sigma_5 \dot{a} + \sigma_6 \right) E_2 \\
  &\mathrm{pr}^2 X (E_2) = \left(\sigma_7 \dot{a} + \sigma_8 \dot{\rho}+ \sigma_9 \right)E_1 + \left(\sigma_{10} \dot{a}^2 + \sigma_{11} \dot{a} +\sigma_{12} \dot{\rho} +\sigma_{13} \dot{N} + \sigma_{14}\dot{P} + \sigma_{15}\right) E_2,
\end{align}
\end{subequations}
the difference being just in the addition of a linear in $\dot{P}$ term on the right hand side of the second equation.

The above system is solved in terms of the functions $\sigma_i$ and the coefficients of the generator $X$ and yields (again we refrain from giving the multipliers $\sigma_i$)
\begin{enumerate}
  \item The parametrization invariance vector $X_{par}$ of \eqref{Xpar}.

  \item The symmetry generator
  \begin{equation} \label{X4}
    X_4 = X_2 + \left[ \left(P+\rho\right) \frac{\partial f}{\partial \rho} - \frac{a}{3} \frac{\partial f}{\partial a} -f(a,\rho) \right] \frac{\partial}{\partial P}
  \end{equation}
  where the $X_2$ is the one of \eqref{X2}, but with the difference that now the $f(a,\rho)$ appearing here is \underline{not} bound by condition \eqref{cond2}; it is an arbitrary function.

  \item The generator
  \begin{equation} \label{X5}
    \begin{split}
      X_5 =& N \left[ \frac{\partial \sigma}{\partial a} - \frac{3 (P+\rho )}{a}\frac{\partial \sigma}{\partial \rho}-\frac{a (\Lambda +8 \pi  \rho ) }{a^2 (\Lambda +8 \pi  \rho )-3 k}\sigma (a,\rho ) \right] \frac{\partial}{\partial N} + \sigma (a,\rho ) \frac{\partial}{\partial a} \\
      & + \left[\frac{3 (P+\rho )^2}{a}\frac{\partial \sigma}{\partial \rho} -(P+\rho ) \frac{\partial \sigma}{\partial a} + \frac{(P+\rho ) }{a}\sigma (a,\rho ) \right] \frac{\partial}{\partial P},
    \end{split}
  \end{equation}
  where $\sigma (a,\rho )$ is another free function.
\end{enumerate}

Thus, we obtain two new point symmetry generators that involve transformations at the level of the pressure $P$ as well. That is two new infinite dimensional symmetry groups characterized by the free functions $f(a,\rho)$ and $\sigma(a,\rho)$. As far as the algebra properties are concerned it is easy to see that $[X_4,X_4]$ is also a $X_4$ vector, $[X_5,X_5]$ belongs to the same class as $X_5$, while $[X_4,X_5]$ gives a composition of the two vectors. In particular, for two vectors $X_4^{(f)}$ and $X_5^{(\sigma)}$ corresponding to some particular $f(a,\rho)$ and $\sigma(a,\rho)$ respectively, we obtain:
\begin{equation*}
  [X_4^{(f)},X_5^{(\sigma)}] = -\left(X_4^{\tilde{(f)}} + X_5^{\tilde{(\sigma)}} \right)
\end{equation*}
where $\tilde{f}(a,\rho)$ and $\tilde{\sigma}(a,\rho)$ are functions related to $f(a,\rho)$ and $\sigma(a,\rho)$ by the differential equations
\begin{equation*}
  \sigma(a,\rho) \frac{\partial f}{\partial a} = \tilde{f}(a,\rho), \quad f(a,\rho) \frac{\partial \sigma}{\partial \rho} = -\tilde{\sigma}(a,\rho) .
\end{equation*}

\section{Relation to previous results}

In this section we explore how our result is connected to what is previously encountered in the literature and how it generalizes some known symmetry properties of the Friedmann equations.

Recently, a new symmetry for the Friedmann equations in the spatially flat FLRW case without cosmological constant and with $P=w \rho$, was reported in \cite{Faraoni}. Let us demonstrate how the latter is a particular case of generator $X_3$ as expressed by \eqref{X3}. The symmetry transformation introduced in \cite{Faraoni} is:
\begin{equation} \label{Farmet}
  \tilde{a} = a^s, \quad \tilde{\rho} = a^{-3 (w+1)(s-1)} \rho, \quad d\tilde{t} = s a^{\frac{3}{2}(w+1)(s-1)} dt,
\end{equation}
with $s=$constant being the parameter of the transformation. The above transformation is mentioned as a symmetry of the Friedmann equations in the cosmological time gauge, where $\tilde{N}=1$. In our language, where we use a generic time gauge, the last of the three relations can be viewed as a change in the lapse function while $t$ itself remains unchanged. In this case this implies that we may take
\begin{equation}
  \tilde{N}= s a^{\frac{3}{2}(w+1)(s-1)} N, \quad \tilde{t}=t.
\end{equation}
With this consideration the symmetry generator of the above transformation reads
\begin{equation*}
  \begin{split}
    \left(\frac{d \tilde{t}}{ds}\Big|_{s=1}\right) \frac{\partial}{\partial t}  + \left( \frac{d \tilde{N}}{ds}\Big|_{s=1}\right) \frac{\partial}{\partial N}+ \left( \frac{d \tilde{a}}{ds}\Big|_{s=1} \right) \frac{\partial}{\partial a} + \left(\frac{d \tilde{\rho}}{ds}\Big|_{s=1} \right) \frac{\partial}{\partial \rho} = \\
    N(1+\frac{3}{2}(w+1)\ln (a)) \frac{\partial}{\partial N} + a\ln (a) \frac{\partial}{\partial a} -3(w+1)\rho \ln (a) \frac{\partial}{\partial \rho}.
  \end{split}
\end{equation*}
But this is exactly the generator you obtain from $X_3$ when you set in \eqref{X3}, $P=w \rho$, $k=\Lambda=0$ and $\xi(a,\rho)= a \ln (a)$.

We can see now how the symmetry, reported in \cite{Faraoni} for the $k=0$ case, can now be extended in a spatially curved universes. By considering $k\neq 0\neq \Lambda$, from \eqref{X3} we derive
\begin{equation}
  X_3 = N\left[1+ \frac{3 \ln (a) \left(k-4 \pi  a^2 \rho  (w+1)\right)}{3 k-a^2 (\Lambda +8 \pi  \rho )}\right]\frac{\partial}{\partial N} + a\ln (a) \frac{\partial}{\partial a} -3(w+1)\rho \ln (a) \frac{\partial}{\partial \rho}
\end{equation}
which leads to the transformation
\begin{equation}
  \tilde{a} = a^s, \quad \tilde{\rho} = a^{-3 (w+1)(s-1)} \rho, \quad \tilde{N}= \frac{s N a^{\frac{1}{2} s (3 w+5)-1} \sqrt{3 k-a^2 (\Lambda +8 \pi  \rho)}}{\sqrt{3 k a^{3 s (w+1)}-8 \pi  \rho a^{2 s+3( w+1)}-\Lambda  a^{s (3 w+5)}}}, \quad \tilde{t}=t .
\end{equation}
Alternatively if we want to consider just invariance starting from the time gauge $\tilde{N}=1$ and considering a time transformation, the above should be interpreted as
\begin{equation}
  \tilde{a} = a^s, \quad \tilde{\rho} = a^{-3 (w+1)(s-1)} \rho, \quad d\tilde{t}= \frac{s a^{\frac{1}{2} s (3 w+5)-1} \sqrt{3 k-a^2 (\Lambda +8 \pi  \rho)}}{\sqrt{3 k a^{3 s (w+1)}-8 \pi  \rho a^{2 s+3( w+1)}-\Lambda  a^{s (3 w+5)}}} dt,
\end{equation}
which is the generalization of \eqref{Farmet} in the case of non zero spatial curvature and cosmological constant.

In \cite{Chimento} another symmetry property of the Friedmann equations in the spatially flat case and without cosmological constant is given. It involves an arbitrary function and thus is related to an infinite dimensional symmetry group. Specifically, it is stated that the transformation
\begin{subequations} \label{chim}
  \begin{align} \label{chim1}
    & \bar{\rho}=\bar{\rho}(\rho) \\ \label{chim2}
    & \bar{H}= \left(\frac{\bar{\rho}}{\rho}\right)^{1/2} H \\ \label{chim3}
    & \bar{P} = -\bar{\rho} + \left(\frac{\rho}{\bar{\rho}}\right)^{1/2} \left( \rho+P\right) \frac{d \bar{\rho}}{d \rho},
  \end{align}
\end{subequations}
where $H= \frac{\dot{a}}{{a}}$ is the Hubble function in the gauge $N=1$ leaves the corresponding equations
\begin{equation} \label{Heqs}
  3H^2- 8 \pi \rho =0 \quad \text{and} \quad \dot{\rho} + 3 H (P+\rho ) =0
\end{equation}
invariant.

Surprisingly, this symmetry transformation does not have an exact equivalent in the study we did in the previous section and this is because the treatment of $H$ as a basic variable changes the formalism. But let us proceed to see in detail what happens and how the aforementioned symmetry can also be recovered.

The generator $X_4$ of \eqref{X4} can induce a change in the pressure like the one implied by \eqref{chim1} if we consider the particular case $k=0=\Lambda$ and $f=f(\rho)$; then,
\begin{equation} \label{X4forchim}
  X_4 = -\frac{N f(\rho )}{2 \rho } \frac{\partial}{\partial N} + f(\rho )\frac{\partial}{\partial \rho} + \left[(P+\rho) f'(\rho )-f(\rho)\right] \frac{\partial}{\partial P}.
\end{equation}
The infinitesimal change implied by the latter up to first order in the parameter of the transformation, say $\lambda$, is
\begin{align} \label{tr1}
  \bar{\rho} & \sim  \rho + \lambda f(\rho ) \\ \label{tr2}
  \bar{N} & \sim N - \lambda \frac{N f(\rho )}{2 \rho } \\ \label{tr3}
  \bar{P} & \sim P + \lambda \left[(P+\rho) f'(\rho )-f(\rho)\right]
\end{align}
If we use \eqref{tr1} and  \eqref{tr2} we may write
\begin{equation}
  \bar{H} = \frac{\dot{\bar{a}}}{{\bar{N}} {\bar{a}}} \sim \frac{1}{N(1-\lambda \frac{f(\rho )}{2 \rho })}\frac{\dot{a}}{a} \sim \left(1+ \lambda \frac{f(\rho)}{2\rho} \right) H
\end{equation}
which is exactly the infinitesimal change emanating from \eqref{chim2} for the $\bar{\rho}(\rho)$ given by \eqref{tr1}, since
\begin{equation}
  \bar{H}= \left(\frac{\bar{\rho}}{\rho}\right)^{1/2} H \sim H \left(1+ \lambda \frac{f(\rho)}{2\rho} \right) .
\end{equation}
So the necessary transformation of $H$ that keeps invariant \eqref{eq1} is attributed to a change of the gauge variable $N$ (or alternatively, as we said before, it can also be seen as a change in $dt$). The situation however for the pressure is not the same: By introducing \eqref{tr1} into \eqref{chim3} we obtain
\begin{equation} \label{chiminfP}
  \bar{P}  = -\bar{\rho} + \left(\frac{\rho}{\bar{\rho}}\right)^{1/2} \left( \rho+P\right) \frac{d \bar{\rho}}{d \rho}\sim P + \lambda \left[(P+\rho) f'(\rho )-f(\rho)\right] + \lambda (P+\rho) f'(\rho)
\end{equation}
which is different from what we see in \eqref{tr3}.

The reason for the above discrepancy rests in the difference of considering $H$ in place of $a$ and $N$, or equivalently $dt$, as the basic variable. The Hubble function in a generic time gauge is $H=\frac{\dot{a}}{N a}$. Any transformation of $H$, when seen at the level of its components, can be attributed to $a$ or to $N$ and $dt$ interchangeably. In the case at hand, of \eqref{X4forchim}, the corresponding change in $H$ is owed to a transformation in $N$, or equivalently the time in the $a$-formalism if we want to keep $N$ fixed (e.g. $N=1$). For maximum compatibility, let us turn to equations \eqref{Heqs} written in the gauge $N=1$ and thus consider that the change in $H$ is owed to a transformation in the time variable. When you perform such a transformation, the $\dot{\rho}$ that we see in the second of \eqref{Heqs} is also going to be transformed. On the other hand, if you forget that $H$ is made up by some specific components and treat it as the basic variable, you have no reason to enforce a transformation of the $\dot{\rho}$ in equation \eqref{Heqs}. Hence, the observed difference between the two situations is passed on to the pressure of each case that now needs to be different in order to keep the equations invariant during these two distinct scenarios. This is the essence of the ``discrepancy" we see between \eqref{chiminfP} and \eqref{tr3}.

As we did in the previous sections, we can proceed and derive the general symmetry generator whose special case is transformation \eqref{chim} in the $H$-formalism of the Friedmann equations. For this to happen we need to consider a generator of the form
\begin{equation}
  X = \chi(t,H,\rho,P) \frac{\partial}{\partial t} +  \xi^1(t,H,\rho,P) \frac{\partial}{\partial H}+  \xi^2(t,H,\rho,P) \frac{\partial}{\partial \rho}+  \xi^3(t,H,\rho,P) \frac{\partial}{\partial P}
\end{equation}
which we require to be a symmetry of the equations
\begin{subequations} \label{eqswithH}
\begin{align} \label{eqH1}
 & E_1^H:= -3 H^2 +8 \pi  \rho + \Lambda =0 \\
 & E_2^H:= 3 H (P+\rho )+ \dot{\rho} =0 \\
 & E_3^H:= -3 H^2-2 \dot{H}+\Lambda -8 \pi  P = 0 .
\end{align}
\end{subequations}
We additionally include the cosmological constant in our considerations.

Since the first equation is algebraic, we start from the last two demanding, i.e.
\begin{align}
  & \mathrm{pr}^{1}X (E_2^H) =0 , \quad \mathrm{mod} \quad E_2^H =0, \quad E_3^H=0, \\
  & \mathrm{pr}^{1}X (E_3^H) =0 , \quad \mathrm{mod} \quad E_2^H =0, \quad E_3^H =0.
\end{align}
The resulting symmetry generator has the additional property $X(E_1)=0$ and it can be split into two parts:
\begin{enumerate}
  \item The first is related to the effect that transformations in the time variable produce to the rest of the quantities and it has the rather complicated form
  \begin{equation} \label{X6}
    \begin{split}
      X_6 =& \chi \frac{\partial}{\partial t} - 4 \pi \int\! (\rho+P) \frac{\partial\chi}{\partial P} dP \; \frac{\partial}{\partial H} - 3 H \int\! (\rho+P) \frac{\partial\chi}{\partial P} dP \; \frac{\partial}{\partial \rho}+ \Bigg[ 3 H \int\! (\rho+P) \frac{\partial\chi}{\partial P}  dP  \\
      &  - 4\pi (\rho+P) \int\! (\rho+P) \frac{\partial^2 \chi}{\partial P\partial H} dP + \int\! (\rho+P) \frac{\partial^2 \chi}{\partial P\partial t}  dP - 3 H (P+\rho ) \int\! (\rho+P)\frac{\partial^2 \chi}{\partial P\partial H} dP \\
      & -(\rho+P) \left(\partial_t\chi+3 H \chi\right) +(\rho+P )^2 \left(3 H \frac{\partial\chi}{\partial \rho} + 4 \pi  \frac{\partial\chi}{\partial H} \right) \Bigg] \frac{\partial}{\partial P},
    \end{split}
  \end{equation}
  where $\chi=\chi(t,H,\rho,P)$ is an arbitrary function.
  \item The second is the vector
  \begin{equation} \label{X7}
    X_7 = h(t,H,\rho)\frac{\partial}{\partial H} + \frac{3 H}{4\pi} h(t,H,\rho)\frac{\partial}{\partial \rho} + \frac{1}{4\pi} \left[(\rho+P ) \left(3 H \frac{\partial h}{\partial \rho}+4 \pi  \frac{\partial h}{\partial H}\right)- \frac{\partial h}{\partial t}- 3 H h(t,H,\rho)\right] \frac{\partial}{\partial P} ,
  \end{equation}
  where $h(t,H,\rho)$ is an arbitrary function of its arguments.
\end{enumerate}

The particular case of the symmetry transformation \eqref{chim} arises when one considers $h(t,H,\rho)= \frac{4 \pi f(\rho)}{3H}$  which yields
\begin{equation} \label{X7chim}
  X_7 = \frac{4 \pi  f(\rho )}{3 H} \frac{\partial}{\partial H} + f(\rho) \frac{\partial}{\partial \rho} + \left[ (P+\rho ) f'(\rho )- f(\rho )- \frac{4 \pi  (P+\rho )}{3 H^2}f(\rho) \right] \frac{\partial}{\partial P} .
\end{equation}
Now this generator is completely compatible with transformation \eqref{chim} in the pressure, with the additional use of the $H^2=\frac{8\pi \rho}{3}$, since the first was written in \cite{Chimento} for the $\Lambda=0$ case. However, simply by mere application of $H^2=\frac{8\pi \rho}{3}+ \frac{\Lambda}{3}$ in \eqref{X7chim} the generalization of \eqref{chim} in the $\Lambda\neq 0$ case can be derived. We thus see how using $H$ as the basic variable leads to different symmetry generators than when applying the theory with the equations expressed in terms of the scale factor.

\section{Generation of new solutions from known ones}

It is widely known that symmetry generators can serve in reducing the order of differential systems of equations. The truth here however is that we deal with a rather simple system. Especially if we use \eqref{eq1} to solve algebraically with respect to the lapse $N$, we need only solve a single first order relation for a given equation of state, i.e. equation \eqref{eq2}. The integrable classes of such relations are well known from the theory of ordinary differential equations. The situation can become challenging if we want to obtain the solution in a particular time gauge (e. g. the cosmic or cosmological time case, $N=1$). Still however, the known cases for which the solution can be derived in terms of known functions have been widely studied. For example, in \cite{Gib1} it is shown that for a spatially non-flat universe with a simple equation of state like $P=w \rho$, only for rational values of $w$ can you write the solution in terms of known functions when $N=1$. A generalization of the previous study for non-linear equations of state can be found in \cite{Gib2}. Thus the main use of generators like those we derived in previous section is into mapping known solutions to new ones. More importantly it is interesting to see what effect the symmetry we use has in the base theory which produces the pressure and the energy density that we study in a model.

As a demonstration and a consistency check, we will start with a few simple examples, where the generators $X_4$ and $X_5$ are used to connect models with different equations of state and hence different cosmological epochs. In the end of the section we will provide a connection of the symmetries we previously derived to the scalar field model and with a simple example we will see how a change based on a symmetry can affect the action principle and the context of the theory.

But let us first start with some generic considerations. The method of generating new solutions from old ones \cite{Stephani,Olver} will be implemented. Let us outline the method.
Suppose the following Lie point symmetry generator is known for some system of partial differential equations
\begin{align}
\mathbf{Z}=\grj^{i}(x^{j},u^{\grb})\frac{\partial }{\partial x^{i}}+\grh^{\gra}(x^{j},u^{\grb})\frac{\partial}{\partial u^{\gra}},\label{eq20}
\end{align}
with $\grj^{i}(x^{j},u^{\grb}),\,\grh^{\gra}(x^{j},u^{\grb})$ known functions. The finite transformation can be obtained through the integral curves of the system
\begin{align}
&\grj^{l}\left[\tilde{x}^{j}(\grl),\tilde{u}^{\grb}(\grl)\right]=\frac{d \tilde{x}^{l}(\grl)}{d \grl},\,\,\grh^{\gra}\left[\tilde{x}^{j}(\grl),\tilde{u}^{\grb}(\grl)\right]=\frac{d \tilde{u}^{\gra}(\grl)}{d \grl},\label{eq7}\\
&\text{with the initial conditions}\nonumber\\
&\tilde{x}^{i}(0)=x^{i},\,\,\tilde{u}^{\gra}(0)=u^{\gra}.\label{eq8}
\end{align}
The result will read in general
\begin{align}
\tilde{x}^{i}=\tilde{x}^{i}(x^{j},u^{\grb};\grl),\,\tilde{u}^{\gra}=\tilde{u}^{\gra}(x^{j},u^{\grb};\grl),\label{eq21}
\end{align}
and since the transformation is invertible,
\begin{align}
x^{i}=x^{i}(\tilde{x}^{j},\tilde{u}^{\grb};\grl),\,u^{\gra}=u^{\gra}(\tilde{x}^{j},\tilde{u}^{\grb};\grl),\label{eq22}
\end{align}
where $\grl$ the parameter.

Some simple special solution of the differential equation will have the form
\begin{align}
u^{\gra}=f^{\gra}(x^{i}),\,\,\text{or},\,\, u^{\gra}-f^{\gra}(x^{i})=0\Leftrightarrow F(u^{\gra},x^{i})=0.\label{eq23}
\end{align}
Hence, the solution is a surface described by \eqref{eq23} in the space with variables $(x^{i},u^{\gra})$. Due to \eqref{eq22} we get
\begin{align}
&F(u^{\gra},x^{i})=0\xRightarrow{\eqref{eq22}}\nonumber\\
&F[x^{i}(\tilde{x}^{j},\tilde{u}^{\grb};\grl),u^{\gra}(\tilde{x}^{j},\tilde{u}^{\grb};\grl)]=0\Rightarrow\nonumber\\
&\tilde{F}(\tilde{x}^{j},\tilde{u}^{\grb};\grl)=0.\label{eq24}
\end{align}
The surface \eqref{eq24} will represent a new form of solution and hopefully we will be able to bring it in the form
\begin{align}
\tilde{u}^{\grb}=\tilde{f}^{\grb}(\tilde{x}^{j};\grl),\label{eq25}
\end{align}
where $\grl$ will act as an integration constant.

When do we expect to find a different form of solution? The surface that represents the original solution should not be \textit{invariant} under the action of the generator, that is
\begin{align}
\mathbf{Z}[F(u^{\gra},x^{i})]\neq{0}.\label{eq26}
\end{align}
If the above does not hold, then we get the same form of solution in the variables $\tilde{x}^{i},\tilde{u}^{\gra}$.

\subsection{Cosmological epochs} \label{exEpochs}

The starting point is the solution of the Einstein's plus perfect fluid equations for flat FLRW $(k=0)$, in the absence of cosmological constant $(\grL=0)$. The assumed equation of state is $(P=w_{1}\grr)$ with $w_{1}<-\frac{1}{3}$, corresponding to the inflation epoch. Furthermore, we are going to choose the gauge $(N=1)$. To this end, the equations become
\begin{align}
&8\grp \grr-3 \left(\frac{\dot{a}}{a}\right)^{2}=0,\label{eqe1}\\
&3(1+w_{1})\frac{\dot{a}}{a}\grr+\dot{\grr}=0,\label{eqe2}
\end{align}
with solutions
\begin{align}
&P=w_{1}\grr,\label{eqe3s1}\\
&N=1,\label{eqe3}\\
&\grr=\grr_{0}a^{-3(1+w_{1})},\label{eqe4}\\
&a(t)=\left[6\grp (1+w_{1})^{2}\grr_{0}\right]^{1/3(1+w_{1})}t^{2/3(1+w_{1})}.\label{eqe4s1}
\end{align}
To achieve our goal we form the generator
\begin{align}
Z=X_{4}+X_{5}.\label{eqe5}
\end{align}
The functions $\grs(a,\grr),f(a,\grr)$ are arbitrary, hence we can choose them properly. We demand the $\partial_{N}$ component of the generator to be equal to zero in order to remain at the same gauge $\tilde{N}=1$, where $\tilde{N}$ the transformed lapse function. This implies that the observers remain co-moving in the transformed system as well. Additionally, we choose the component $\partial_{\grr}$ to be equal to $\grr$, which also leads the component of $\partial_{P}$ to be independent of the scale factor $a$. Thus, the transformed equation of state will remain only a function of $\tilde{\grr},\tilde{P}$. These requirements reduce the form of the functions and the generator to
\begin{align}
&\grs(a,\grr)=\frac{a}{2}\ln a,\,f(a,\grr)=\grr,\label{eqe6}\\
&Z=\frac{a}{2}\ln a\partial_{a}+\grr \partial_{\grr}+\frac{P-\grr}{2}\partial_{P}.\label{eqe7}
\end{align}
The equations \eqref{eq7} for the generator $Z$ read
\begin{align}
\frac{d\tilde{t}}{d\lambda}\Big|_{\lambda=0}=0,\quad \frac{d\tilde{N}}{d\lambda}\Big|_{\lambda=0} =0,\quad \frac{d\tilde{a}}{d\lambda}\Big|_{\lambda=0}=\frac{\tilde{a}}{2}\ln \tilde{a}, \quad \frac{d\tilde{\rho}}{d\lambda}\Big|_{\lambda=0}=\tilde{\rho},\quad \frac{d\tilde{P}}{d\lambda}\Big|_{\lambda=0}=\frac{\tilde{P}-\tilde{\grr}}{2},\label{eqe8}
\end{align}
where $\grl$ the parameter of the curves, chosen so that at $\grl=0$ the transformation is the identity. The transformation as well as the inverse are
\begin{align}
&\tilde{t}=t,\,\tilde{N}=N,\,\tilde{a}=a^{e^{\grl/2}},\,\tilde{\grr}=e^{\grl}\grr,\,\tilde{P}=-e^{\grl}\grr+e^{\grl/2}\left(P+\grr\right),\label{eqe9}\\
&t=\tilde{t},\,N=\tilde{N},\,a=\tilde{a}^{e^{-\grl/2}},\,\grr=e^{-\grl}\tilde{\grr},\,P=e^{-\grl}\left[-\tilde{\grr}+e^{\grl/2}\left(\tilde{P}+\tilde{\grr}\right)\right].\label{eqe10}
\end{align}
All we have to do now is to use \eqref{eqe10} in \eqref{eqe3s1}-\eqref{eqe4s1} and solve with respect to the new variables
\begin{align}
&\tilde{P}=\left(-1+(1+w_{1})e^{-\grl/2}\right)\tilde{\grr},\label{eqe11}\\
&\tilde{N}=1,\label{eqe12}\\
&\tilde{\grr}=e^{\grl}\grr_{0}\tilde{a}^{-3(1+w_{1})e^{-\grl/2}},\label{eqe13}\\
&\tilde{a}(\tilde{t})=\left[6\grp (1+w_{1})^{2}\grr_{0}\right]^{1/3(1+w_{1})e^{\grl/2}}\tilde{t^{2/3(1+w_{1})e^{\grl/2}}}.\label{eqe14}
\end{align}
By the choice $\grl=2\ln\frac{3(1+w_{1})}{4}$ and the redefinition $\grr_{0}=\frac{16}{9(1+w_{1})^{2}}\tilde{\grr}_{0}$, the inflation transforms to the radiation epoch solution.
\begin{align}
&\tilde{P}=\frac{1}{3}\tilde{\grr},\label{eqe20}\\
&\tilde{N}=1,\label{eqe21}\\
&\tilde{\grr}=\tilde{\grr}_{0}\tilde{a}^{-4},\label{eqe22}\\
&\tilde{a}(\tilde{t})=2\left(\frac{2\grp}{3}\tilde{\grr}_{0}\right)^{1/4}\tilde{t}^{1/2}.\label{eqe23}
\end{align}
Now we can repeat the transformation \eqref{eqe10} with a different parameter $\grl$ and as known solutions the \eqref{eqe20}-\eqref{eqe23}. There exist a parameter $\grl$ which can transform the radiation epoch solution into the dust epoch. In the following table we present the values of the parameters $\grl$ and the corresponding epoch's, as well as the equations of state.

\begin{table}[h]
\centering
\begin{tabular}{ |c|c|c| }
\hline
\textbf{Transitions} &  $\grl$ & Equations of state \\
\hline
Inflation$\rightarrow$ Radiation, & $2\ln\frac{3(1+w_{1})}{4}$ & $w_{i}<-\frac{1}{3}\rightarrow w_{r}=\frac{1}{3}$\\
\hline
Radiation$\rightarrow$ Dust, & $\ln\frac{16}{9}$ & $w_{r}=\frac{1}{3}\rightarrow w_{d}=0$\\
\hline
Dust$\rightarrow$ De Sitter, & $\rightarrow \infty$ & $w_{d}=0\rightarrow w_{dS}=-1$\\
\hline
\end{tabular} \caption{Mappings between solutions for various values of $\lambda$.} \label{tab1}
\end{table}

As we can see, the last transition can be obtained only as a limit of $\grl$.

\subsection{Stiff plus generalized (Anti-) Chaplygin fluid from stiff equation of state} \label{exLCDMtoChap}

Let us assume an equation of state describing stiff matter $P=\grr$ and $k=0$ with arbitrary $\grL$. The solution is
\begin{align}
&P=\grr,\label{sm1}\\
&N=1,\label{sm2}\\
&\grr=\grr_{0}a^{-6}\label{sm3}\\
&a(t)=\sqrt{2}\left(\frac{\grp \grr_{0}}{\grL}\right)^{1/6}\sinh^{1/3}\left(\sqrt{3\grL}t\right).\label{sm4}
\end{align}
For this case, only the $X_{4}$ generator will be used, with the function $f(a,\grr)$ chosen to have the form
\begin{align}
&f(a,\grr)=\grr^{-\grn},\label{sm5}\\
&X_{4}=-\frac{4\grp N \grr^{-\grn}}{\grL+8\grp \grr}\partial_{N}+\grr^{-\grn}\partial_{\grr}-\grr^{-1-\grn}\left[\grn P+\left(1+\grn\right)\grr\right]\partial_{P},\label{sm6}
\end{align}
where $\grn$ some arbitrary constant. The transformed solutions read
\begin{align}
&\tilde{P}=\tilde{\grr}-\frac{2\grl(1+\grn)}{\tilde{\grr}^\nu},\label{sm7}\\
&\tilde{N}=\frac{\sqrt{\grL+8\grp \tilde{\grr}}}{\sqrt{\grL+8\grp\left[\left(1+\grn\right)\grl+\tilde{\grr}^{1+\grn}\right]^{\frac{1}{1+\grn}}}}
,\label{sm8}\\
&\tilde{\grr}=\left[\left(\grr_{0}\tilde{a}^{-6}\right)^{1+\grn}+ \left(1+\grn\right)\grl\right]^{1/(1+\grn)},\label{sm9}\\
&\tilde{a}=\sqrt{2}\left(\frac{\grp \grr_{0}}{\grL}\right)^{1/6}\sinh^{1/3}\left(\sqrt{3\grL}\tilde{t}\right).\label{sm10}
\end{align}
From \eqref{sm7} the generalized Anti-Chaplygin matter can be obtained if $\lambda<0$ while the Chaplygin for $\lambda>0$ (assuming of course for the power of $\tilde{\rho}$ in the second term $\nu>0$). Notice however, that in this case the observers are not co-moving any more since there is a lapse function $\tilde{N}\neq{1}$.

\subsection{Scale-factor dependent equation of state from dust solution} \label{example3}

The only difference with the first subsection is to choose different functions $\grs(a,\grr),f(a,\grr)$ and our starting point is the dust solution corresponding to $P=0$. Again, we want to maintain the lapse gauge but there is no other restriction. The functions and the generator are
\begin{align}
&\grs(a,\grr)=\gra^{2},\,f(a,\grr)=2a \grr,\label{eqe24}\\
&Z=\gra^{2}\partial_{a}+2a \grr\partial_{\grr}+a\left(P-\frac{5}{3}\grr\right)\partial_{P}.\label{eqe25}
\end{align}
The induced transformation reads
\begin{align}
t=\tilde{t},\,N=\tilde{N},\,a=\frac{\tilde{a}}{1+\grl\tilde{a}},\,\grr=\frac{\tilde{\grr}}{(1+\grl \tilde{a})^{2}},\,P=\frac{3\tilde{P}(1+\grl \tilde{a})+5\grl \tilde{a}\tilde{\grr}}{3(1+\grl \tilde{a})^{2}},\label{eqe26}
\end{align}
and the new solution are
\begin{align}
&\tilde{P}=-\frac{5\grl \tilde{a}}{3(1+\grl \tilde{a})}\tilde{\grr},\label{eqe27}\\
&\tilde{N}=1,\label{eqe28}\\
&\tilde{\grr}=\frac{(1+\grl \tilde{a})^{5}}{\tilde{a}^{3}}\grr_{0},\label{eqe29}\\
&\tilde{a}(\tilde{t})=\frac{(6\grp \grr_{0})^{1/3} \tilde{t}^{2/3}}{1-\grl (6\grp \grr_{0})^{1/3} \tilde{t}^{2/3}}.\label{eqe30}
\end{align}
As we can observe from \eqref{eqe27} the equation of state depends on the scale factor. Furthermore, by use of \eqref{eqe30} and the redefinition of the parameter $\grl$ as follows $\grl=-\frac{3^{2/3}}{5(2\grp \grr_{0})^{1/3}}w$ the equation of state may be written as
\begin{align}
\tilde{P}=w \tilde{t}^{2/3}\tilde{\grr}.\label{eqe31}
\end{align}
The equation of state is time dependent and this dependence is similar to the dependence of the original's solution, scale factor.

The infinite symmetry freedom offers infinite possibilities. The ability to obtain different equations of state from the simplest original one $(P=0)$ is due to the assumption that $\grr$ and $P$ are independent variables.

\subsection{Effects of symmetries on the base theory}

With generators like $X_i$, $i=1,2,3$ we derived symmetry transformations that keep you in the same theory, in the sense that the equation of state remains invariant. On the other hand, transformations derived from $X_i$, $i=4,...,7$ provide us with a connection between different types of fluids. The first set can mostly serve in the reduction of the order of equations, although due to the simplicity of the latter, such reductions are expected to fall into known integrable classes. Alternatively these generators can in principle be used to derive a more general solution from a partial solution which is known e.g. a solution for a specific value of the integration constants. The significance of the second set of generators however is much more ample since they are able to provide a correspondence between different models. More importantly we can derive information over the base theory as the effect of such a transformation. In the literature there have been interesting cases where new solutions are derived from old through the application of transformations \cite{newold}.

Let us take as an example generator $X_7$ from \eqref{X7} which is a symmetry of \eqref{eqswithH}. Consider that we have in our hands a given model with a Hubble function $H(t)$ and assume that we want to study which model would result in a new Hubble expansion rate $H\mapsto \bar{H}= H^{\lambda}$. We know that a power law transformation in the $H$ variable can be generated by utilizing  in \eqref{X7} a function $h(t,H,\rho)= H \ln H$. With such a choice the generator $X_7$ is written as
\begin{equation} \label{X7example}
  X_7 = H \ln H \frac{\partial}{\partial H} + \frac{3 }{4 \pi } H^2 \ln H \frac{\partial}{\partial \rho} + \left[ \left(-\frac{3 H^2}{4 \pi }+P+\rho \right) \ln H + P+ \rho \right] \frac{\partial}{\partial P},
\end{equation}
which straightforwardly provides us with the necessary transformation law in $\rho\mapsto \bar{\rho}$ and $P\mapsto \bar{P}$. In total we obtain:
\begin{subequations}
\begin{align} \label{Hbar}
  H & = \bar{H}^{\frac{1}{\lambda}}, \\ \label{rHtr}
  \rho & = \bar{\rho}  + \frac{3}{8\pi} \left(\bar{H}^{\frac{2}{\lambda}} - \bar{H}^2 \right), \\ \label{PHtr}
  P & = \frac{\bar{H}^{\frac{1-\lambda}{\lambda}}}{\lambda} \left(\bar{P} + \bar{\rho} \right) - \bar{\rho} - \frac{3}{8\pi} \left(\bar{H}^{\frac{2}{\lambda}} - \bar{H}^2 \right) ,
\end{align}
\end{subequations}
where $\lambda$ is the parameter of the transformation; the identity mapping is obtained for $\lambda=1$.

Note that the above transformation is generic, i.e. it does not depend on a specific initial model. Our starting assumption is that whatever $H$ we have, we want to find which theory would reproduce an $H^\lambda$ expansion rate. In order to find the latter we need to set some specific initial state. For example if we start from a theory with $P=w \rho$, then with the use of \eqref{rHtr}, \eqref{PHtr} and the constraint equation $-3 \bar{H}^2 +8 \pi  \bar{\rho} + \Lambda =0$ to substitute $\bar{H}$ from, we can easily find that the new equation of state reads
\begin{equation}
\bar{P}(\bar\rho) = \frac{3^{-\frac{\lambda +1}{2 \lambda }} \lambda  (w+1)}{8 \pi} (\Lambda +8 \pi  \bar{\rho})^{\frac{\lambda -1}{2 \lambda }} \left(3 (\Lambda +8 \pi \bar{\rho})^{\frac{1}{\lambda} }-3^{\frac{1}{\lambda} } \Lambda \right) - \bar{\rho} .
\end{equation}
It is true that in this form not a lot can be said about the theory that produces such a complicated equation of state. But we must keep in mind the various different matter contents may produce the effective fluids' pressure and energy density that we see in Einstein's equations.

To get an example of such a situation lets consider that the base theory producing the fluid is a single scalar field which is minimally coupled to gravity. We have the well known relations
\begin{equation} \label{Prtophi}
  P = \frac{\dot{\phi}}{2} - V(\phi) , \quad \rho = \frac{\dot{\phi}}{2} + V(\phi),
\end{equation}
for the pressure and energy density of the effective fluid. The above expressions give a transformation rule that can translate the symmetry vector $X_7$ we previously used to new variables $(t,H,\rho,P) \mapsto (t,H,\dot{\phi}, V)$. Note that the pressence of $\dot{\phi}$ in the coordinates does not make this a higher order symmetry in the new variables because $V$ is considered independent of $\dot{\phi}$ (and it is so as long as we do not choose a specific $V(\phi)$). In the new coordinates, the vector $X_7$ from \eqref{X7example} reads
\begin{equation}
  X_7 = H \ln H \frac{\partial}{\partial H} + \frac{1}{2} (\ln H+1) \dot{\phi} \frac{\partial}{\partial \dot{\phi}} + \left[ \frac{1}{4 \pi } \left(3 H^2-2 \pi  \dot{\phi}^2\right) \ln H - \frac{\dot{\phi}^2}{2} \right] \frac{\partial}{\partial V},
\end{equation}
and it can be easily verified that it is a Lie-point symmetry of equations \eqref{eqswithH} when \eqref{Prtophi} are used (the same of course can be checked for the transformed version of the generic vector \eqref{X7}).

The induced transformation $(\dot{\phi},V)\mapsto (\dot{\psi},U)$ to a new scalar field model, apart from the \eqref{Hbar} which remains unchanged, is
\begin{subequations}\label{psiU}
\begin{align}
  \dot{\phi} & = \sqrt{\frac{1}{\lambda }} \bar{H}^{\frac{1}{2} \left(\frac{1}{\lambda }-1\right)}  \dot{\psi}, \\
  V & = U + \frac{3}{8 \pi } \left(\bar{H}^{\frac{2}{\lambda}}-\bar{H}^2\right) + \frac{ \left(\lambda \bar{H} -\bar{H}^{1/\lambda }\right)}{2 G \lambda } \dot{\psi}^2.
\end{align}
\end{subequations}
Again our objective is to see what theory, i.e. which scalar field potential $U(\psi)$ can change our Hubble expansion rate from $H$ (which is given by a $V(\phi)$) to $H^\lambda$. Of course we need to set our initial point through a starting $H$. To achieve this lets take as an example the well known exact solution of an exponential potential when the cosmological constant is zero, i.e. from now on we assume $\Lambda=0$
\begin{equation} \label{solexp}
  H(t) = \frac{\kappa}{t}, \quad \phi(t)= -\frac{\sqrt{\kappa }}{2 \sqrt{\pi }} \ln t, \quad V(\phi) = \frac{\kappa  (3 \kappa -1) }{8 \pi }e^{\frac{4 \sqrt{\pi } \phi}{\sqrt{\kappa }}} .
\end{equation}
Please notice that this is a partial solution, the full general solution of an exponential scalar field can be derived in a rather complicated form \cite{Russo,scexpo}. However, the partial solution \eqref{solexp} is sufficient for the needs of our example and apart from being simpler it also corresponds to a linear equation of state $P = \left(\frac{2}{3 \kappa }-1\right) \rho$.

By using the known solution \eqref{solexp} it is easy to derive from \eqref{Hbar} and \eqref{psiU}, that in order to obtain an expansion rate $\bar{H} = H^\lambda =\frac{\kappa^\lambda}{t^\lambda} $ we need a potential of the form
\begin{equation}
  U(\psi) = \frac{3}{8} \pi ^{\frac{\lambda +1}{\lambda -1}} \left(\frac{(\lambda -1)^4}{\kappa^2  \lambda^2 }\right)^{\frac{\lambda }{\lambda -1}} \psi^{\frac{4 \lambda }{\lambda -1}} - \frac{\lambda^2}{8} \pi ^{\frac{2}{\lambda -1}} (\lambda -1)^{\frac{2 (\lambda +1)}{\lambda -1}}  \left(\frac{1}{\kappa^2  \lambda^2 }\right)^{\frac{\lambda }{\lambda -1}} \psi^{\frac{2 (\lambda +1)}{\lambda -1}}
\end{equation}
while the scalar field is
\begin{equation}
  \psi(t) = \frac{\sqrt{\lambda } \kappa^{\lambda /2} }{\sqrt{\pi } (\lambda -1)}t^{\frac{1}{2}-\frac{\lambda }{2}}
\end{equation}
For the values $0<\lambda<1$ this type of potential $U(\psi)$ is said to describe an ``intermediate" inflation \cite{BarrowS}, in the sense that the expansion rate is something between a power law and an exponential; the solution is extensively studied in \cite{BarrowS,BarrowL}. We just note, that in the parametrization we use here the limit $\lambda>>1$, $\kappa \sim \lambda$ reproduces a Higgs type of scalar field potential. In addition, if you allow $\lambda$ to be negative, then the solution corresponds to a phantom field with $\psi(t)$ becoming imaginary.

This is how a symmetry transformation may help us to obtain in an algorithmic manner the background theory that we would need in order to change the expansion rate in a desired way. We demonstrated it with a simple example, however as we see by the form of the generators, one can choose more complicated combinations among the functions involved in a model. The symmetry generators, depending on the change we want to enforce in some of the observed parameters, can lead us to certain modifications in the action of the underlying theory which we need to study.

The above process for which we got the corresponding generator expressed in term of the scalar field potential and velocity can also be applied with the other generators expressed in the $(t,N,a,\rho,P)$ variables. But in this case the parametrization invariant expressions, $P = \frac{\dot{\phi}}{2N} - V(\phi)$ and $\rho = \frac{\dot{\phi}}{2N} + V(\phi)$ need to be used instead of \eqref{Prtophi}.

\section{Conclusion}

Symmetries of differential equations are of wide interest since they provide an effective method of simplifying and solving a given set of equations. Among other things, they can be used to reduce the order of the equations, find symmetry invariant solutions or utilize already known solutions to acquire new ones.

Given the continually rising interest in cosmological solutions, especially in what regards the nature of dark energy, we thoroughly studied the point symmetries of the Friedmann equations. Due to the fact that the system basically comprises of first order equations, namely \eqref{eq1} and \eqref{eq2}, the resulting symmetry groups are infinite dimensional. In order to be as generic as possible we considered the equations in their original form, prior to assuming any gauge fixing condition.

We separately examined two main scenarios: In the first case a specific equation of state is given and substituted inside the equations of motion. As a result the emanating symmetry vector has the property of leaving the equation of state invariant as well. The second, more intriguing, possibility involved taking the pressure as being a separate dynamical variable which is allowed to take part in the transformation. Performing such a change of variables can lead to a set of Friedmann equations corresponding to a fluid governed by a different equation of state than that of the initial system.

In the first scenario - where we  assumed some ab initio given equation of state - we derived the general symmetry vectors corresponding to dependences of the form $P=P(a,\rho)$; plus an additional possibility where we considered dependence from some power of the first derivative of the scale factor $a$. To the above general forms of equations of state there corresponds a wide class of fluids that can be considered in various cosmological configurations.

However, the most interesting situation emerges when the pressure is allowed to be affected by the transformation. This leads to a distinct class of symmetry vectors which can be used to map solutions among models with different equations of state. Again infinite dimensional symmetry groups are involved, whose generators we derived. To demonstrate their possible applications, we used a transformation to ``scan" through the solutions of the various epochs of the universe (see the example of section \ref{exEpochs}). We also started from a simple solution corresponding to a typical linear equation of state, like $P= \rho$, and showed how - through an appropriate symmetry transformation - it can be turned into the corresponding solution of a more complicated type of fluid of stiff matter plus Chaplygin gas (see section \ref{exLCDMtoChap}). We even used this approach to pass into a solution linked to a time dependent equation of state parameter (see example \ref{example3}). Considering that the aforementioned symmetry groups are infinite dimensional there is an immensely large set of possibilities to be explored in this respect, where simple solutions can be used to derive those of more perplexing configurations.

In the case of spatially flat space-times, $k=0$, where the equations can be formulated purely in terms of the Hubble parameter $H$, we observed that the symmetry generator changes when $H$ is to be considered as the basic variable of the transformation in place of the scale factor $a$. As we saw this genuinely results into distinct symmetry transformations.

To underline the generalization that our work offers, we compared with previously known results of symmetry transformations of the Friedmann equations. We showed how these known symmetries emerge as special cases from the generators we derived and how they can be extended to a wider class of cases (e.g. non-vanishing k and/or $\Lambda$) since they are part of more general transformations. Additionally, we considered the connection to a single scalar field theory and the effective fluid that it creates. Through a simple example we demonstrated how one can induce transformations based on a desired expression for an observable quantity, like the Hubble function, to obtain information on the underlying theory that is needed to produce it as a result. It is quite interesting to note at this point, that the Friedmann equations are seen to have a wider spectre than cosmology. There are various recent works that relates them to other interesting physical problems \cite{Far1,Far2}. Thus the study of their symmetry groups is expected to have a wide application.

Lastly, it is noteworthy to mention that even larger symmetry groups for the system under consideration can be achieved: If one allows $N$ to be simply defined by the quadratic constraint $E_2$, then the initial vector will not have a $\frac{\partial}{\partial N}$ component and it would only have to satisfy one symmetry   condition, namely that  for $E_1$. The application of this idea to this, as well as other cosmological systems, is currently under investigation.

\bigskip%
\bigskip

{{\textbf{Acknowledgements:}}} AP acknowledges financial support of
Agencia Nacional de Investigaci\'{o}n y Desarrollo - ANID through the
program FONDECYT Iniciaci\'{o}n grant no. 11180126. Additionally, by
Vicerrector\'{\i}a de Investigaci\'{o}n y Desarrollo Tecnol\'{o}gico at
Universidad Catolica del Norte.

\end{document}